\documentclass{JHEP3} % 10pt is ignored!

\usepackage{epsfig,multicol}

\title{Hawking radiation from a Vaidya black hole: a semi-classical approach and beyond}

\author{Haryanto M. Siahaan$^a$ and Triyanta$^b$\\
\textit{Theoretical High Energy Physics and Instrumentation Division,\\
Faculty of Mathematics and Natural Sciences, Institut Teknologi Bandung,\\
Jalan Ganesha 10, Bandung 40132, Indonesia}.\\
$^a$\email{anto\_102@students.itb.ac.id}\\
$^b$\email{triyanta@fi.itb.ac.id}}

\received{\today} 		%%
\accepted{\today}		%% These are for published papers.

\preprint{arXiv:0811.1132 [gr-qc]}	% OR: \preprint{Aaaa/Mm/Yy\\Aaa-aa/Nnnnnn}
\abstract{We derive the Hawking radiation for Vaidya black hole in the tunneling picture from the corresponding single particle action by the use of the radial null geodesic and the Hamilton-Jacobi method (beyond semi-classical approximation). Both  results are then analyzed and compared.}


\begin{document} 

\section{Introduction}
%Present an introduction to your paper here.
In 1974, Hawking startled the physics community by proving that black hole evaporates particles \cite{1}. It contradicts with classical general relativistic definition of a black hole, an object that nothing can escape from it \cite{2}, \cite{3}. Hawking's derivation was a quantum field theoretically. The semi-classical approach in modeling the black hole radiation as a tunneling effect was developed in the past decade \cite{4}, \cite{5}, \cite{6}, \cite{7} and attracted many attentions [e.g. \cite{16}]. This is mostly because of its bold physical picture and it is easier, mathematically. \newline
There are two ways to perform semi-classical analysis for a black hole radiation. The first is by the use of radial null geodesic method developed by Parikh and Wilczek \cite{3}, \cite{4}. In this method, one has to get the expression dr/dt from the radial null geodesic condition, $
ds^2  = d\Omega  = 0
$, for a metric that has the form $
ds^2  =  - a\left( r \right)dt^2  + b\left( r \right)dr^2  + r^2 d\Omega ^2 
$. Then, the obtained expression is used to calculate the imaginary part of the action. We then end up with the obtaining of the Hawking temperature $
T_H  = \left( {k\beta } \right)^{ - 1} 
$, after equating the transition rate for a particle in a tunneling process $
\Gamma  \sim \exp \left[ {2{\mathop{\rm Im}\nolimits} S} \right]
$ with the Boltzmann factor $
\exp \left[ { - \beta E} \right]
$. \newline
The second method was developed by Padmanabhan et al. \cite{5} and it attracted some applications (e.g. in 2-d stringy black holes \cite{17}). In the method, the scalar wave function is determined by the ansatz $
\phi \left( {r,t} \right) = \exp \left[ {{{ - iS\left( {r,t} \right)} \mathord{\left/
 {\vphantom {{ - iS\left( {r,t} \right)} \hbar }} \right.
 \kern-\nulldelimiterspace} \hbar }} \right]
$ where $S\left( {r,t} \right)
$ is the action for a single scalar particle. Inserting this ansatz into the Klein-Gordon equation in a gravitational background, one yields an equation for the action $S\left( {r,t} \right)
$ which can be solved by the Hamilton-Jacobi method. After obtaining the action, one can get the probability for outgoing and ingoing particles, $P_{out} = \left| {\phi _{{\rm{out}}} } \right|^2 
$ and $P_{in} = \left| {\phi _{{\rm{in}}} } \right|^2 
$, respectively. The Hawking temperature can be obtained by using the 'principle of detailed balance' [5], $P_{out} = \exp \left[ { - \beta E} \right]P_{in} = \exp \left[ { - \beta E} \right]
$, since all particles must be absorbed by the black hole.\newline
An interesting improvement has recently been made by Banerjee and Majhi \cite{8}. They expand the action   for single particle in the power series of Planck constant, $S\left( {r,t} \right) = \sum\nolimits_{n = 0} {\alpha _n \hbar ^n S_n \left( {r,t} \right)} 
$. By considering that, for all $n$, $S_n \left( {r,t} \right)
$ and $S_{n + 1} \left( {r,t} \right)
$ are proportional to each other with the same proportionality, one can write $S\left( {r,t} \right) = \left( {1 + \sum\nolimits_{n = 1} {\alpha _n \hbar ^n } } \right)S_0 \left( {r,t} \right)
$. By using the method of Padmanabhan et al., the action gives the correction value of the Hawking temperature. Even though the correction is in principle negligible, due to the very small of the Planck constant, one could regard this as an effect of quantum analysis in the semi-classical quantum gravity. Several developments have been made, including fermionic consideration and the relation of the method to trace anomaly \cite{15}.\newline
In this paper, we consider a more general metric with the mass of black hole is time ($t$) and radius ($r$) dependent. A semiclassical approach for a dynamical black hole has also recently worked in \cite{18}. As generally known, this subject is well described by the Vaidya metric. To get an exact ($t-r$) dependence of the Hawking temperature, we insist to work in using Schwarzschild like metric for the Vaidya space time \cite{9} and impose the condition for a very slowly varying mass of the black hole, $\left| {\partial m/\partial r} \right| \equiv \left| {m'} \right| <  < 1
$ and $\left| {\partial m/\partial t} \right| \equiv \left| {\dot m} \right| <  < 1
$. In this approximation, the squared $m'$ and $\dot m
$ are well approximated to zero. This consideration will be useful later in showing the proportionality between each term of action's expansion. Our aim is to see the ($t-r$) dependence of Hawking temperature in the case of varying mass of black hole. We derive it both in the radial null geodesic and Hamilton-Jacobi method. For Hamilton-Jacobi method, we directly consider all of the expansion terms of action (beyond the method of Padmanabhan et al.) as Banerjee and Majhi had done for some metrics with constant black hole masses.\newline  
The organization of our paper is as follows. In the second section, we will derive the Hawking temperature for a metric with a varying mass by the use of radial null geodesic method. In the third section, the Hawking temperature is obtained by the Hamilton Jacobi method, by considering all terms in action expansion. In the last section, we give a conclusion for our work. For the rest of this paper, we use the unit dimension: Newton constant, light velocity in vacuum, and Boltzman constant, $G = c = k_B  = 1
$.
\section{Hawking Temperature from Radial Null Geodesic Method in Vaidya Space-Time}
We start with the metric derived by Farley and D'Eath \cite{9} for Vaidya space-time
\begin{eqnarray}
	ds^2  =  - \left( {\frac{{\dot m}}{{x\left( m \right)}}} \right)^2 \left( {1 - \frac{{2m}}{r}} \right)dt^2  + \left( {1 - \frac{{2m}}{r}} \right)^{ - 1} dr^2  + r^2 d\Omega ^2 .\label{eq:1}
\end{eqnarray}
In the above, the black hole mass $m$ varies with time $t$ and radius $r$, $m \equiv m\left( {r,t} \right)
$. $x\left( m \right)
$ is an arbitrary function of mass and $d\Omega ^2  = d\theta ^2  + \sin ^2 \theta d\varphi ^2 
$ is the metric of 2-sphere. Later, rather than adopting the law of mass evolution by Hawking \cite{10}, \cite{11}, $dm/dt =  - C/m^2 
$, we choose a weaker condition, $\partial m/\partial t \equiv \dot m =  - C(m)/m^2 
$, which is a partial derivative, in contrast to the Hawking law that is an exact differentiation. It will be shown that this choice leads to a better form of Hawking temperature. The variable $C\left( m \right)
$ in this expression counts the number of particles emitted by the black hole with mass $m$. This variable will increase for decreasing mass \cite{9} and for early stage of evaporation, we could consider that this variable is a small valued-function because the evaporated mass is small compared to the total mass. In this line of thought, putting $x\left( m \right) =  - C(m)/m
$ is still valid, at least at the early stage of evaporation. Our purpose in setting $x\left( m \right)
$ in this form is to reduce the variable contained in the metric and later on, for the case of fixed mass with respect to radius, to obtain time dependent Hawking temperature \cite{11}, $T_H \left( t \right) = \hbar /8\pi m(t)
$.\newline 
The above considerations lead the metric (\ref{eq:1}) into the form 
\begin{eqnarray}
	ds^2  =  - \left( {\frac{1}{{m}}} \right)^2 \left( {1 - \frac{{2m}}{r}} \right)dt^2  + \left( {1 - \frac{{2m}}{r}} \right)^{ - 1} dr^2  + r^2 d\Omega ^2 . \label{eq:2}
\end{eqnarray}
It has the general form
\begin{eqnarray}
	ds^2  =  - F\left( {r,t} \right)dt^2  + G\left( {r,t} \right)^{ - 1} dr^2  + r^2 d\Omega ^2 .\label{eq:3}
\end{eqnarray}
where in our case $F\left( {r,t} \right) \equiv  m^{ - 2} \left( {1 - 2mr^{ - 1} } \right)
$ and $G\left( {r,t} \right) \equiv 1 - 2mr^{ - 1} 
$. The general form (\ref{eq:3}) will simplify our next calculation. Unless for specific purposes, we will write $F\left( {r,t} \right)
$ and $G\left( {r,t} \right)
$ as $F$ and $G$, respectively, for the sake brevity.\newline
It turns out that the metric (\ref{eq:2}) has a coordinate singularity at $r_h  = 2m
$ which of course is time dependent. Painleve transformation which is used to remove coordinate singularity for a metric with time-like Killing vector, is also applicable in this analysis. Transforming 
\begin{eqnarray}
	dt \to dt - \sqrt {\frac{{1 - G}}{{FG}}} dr \label{eq:4}
\end{eqnarray}
the metric (\ref{eq:3}) changes into
\begin{eqnarray}
	ds^2  =  - F\left( {r,t} \right)dt^2  + 2F\sqrt {\frac{{1 - G}}{{FG}}} dtdr + dr^2  + r^2 d\Omega ^2 ,\label{eq:5}
\end{eqnarray}
and therefore no coordinate singularity is found. Thus, by such a Painleve transformation, it understandable that, in principle, a coordinate singularity in general relativity can be removed only by changing coordinate, without defining new physical condition or theorem. In the consideration that a tunneled particle moves on the path with no singularity, (\ref{eq:5}) will be useful in describing particle's dynamics. For radial null geodesic, $ds^2  = d\Omega ^2  = 0
$, the differentiation of radius with respect to time can be obtained from (\ref{eq:5}) as
\begin{eqnarray}
	\frac{{dr}}{{dt}} = \sqrt {\frac{F}{G}} \left( { \pm 1 - \sqrt {1 - G} } \right),\label{eq:6} 
\end{eqnarray}
where $+ ( - )
$ signs denote outgoing(ingoing) radial null geodesics.\newline
Near the horizon, we can expand the coefficient $F$ and $G$ by the use of Taylor expansion. Since $F$ and $G$ are $\left( {t - r} \right)
$ dependent, and we only need their approximation values for short distances from a point (horizon), we could apply the Taylor expansion at a fixed time. So, we can write
\begin{eqnarray}
	\left. {F\left( {r,t} \right)} \right|_t  \simeq \left. {F'\left( {r,t} \right)} \right|_t \left( {r - r_h } \right) + \left. {O\left( {\left( {r - r_h } \right)^2 } \right)} \right|_t ,\label{eq:7}
\end{eqnarray}
and
\begin{eqnarray}
	\left. {G\left( {r,t} \right)} \right|_t  \simeq \left. {G'\left( {r_h ,t} \right)} \right|_t \left( {r - r_h } \right) + \left. {O\left( {\left( {r - r_h } \right)^2 } \right)} \right|_t .\label{eq:8}
\end{eqnarray}
By the approximations (\ref{eq:7}) and (\ref{eq:8}) above, the dependence of radius to time in (\ref{eq:6}) can be approached by
\begin{eqnarray}
	\frac{{dr}}{{dt}} \simeq \frac{1}{2}\sqrt {F'\left( {r_h ,t} \right)G'\left( {r_h ,t} \right)} \left( {r - r_h } \right) .\label{eq:9}
\end{eqnarray}
Now, we discuss the action of outgoing particle through the horizon. In the original work by Parikh and Wilczek \cite{3}, the imaginary action is written as
\begin{eqnarray}
	{\mathop{\rm Im}\nolimits} S = {\mathop{\rm Im}\nolimits} \int\limits_{r_{{\rm{in}}} }^{r_{{\rm{out}}} } {p_r dr}  = {\mathop{\rm Im}\nolimits} \int\limits_{r_{{\rm{in}}} }^{r_{{\rm{out}}} } {\int\limits_0^{p_r } {dp_r 'dr} } = {\mathop{\rm Im}\nolimits} \int\limits_{r_{{\rm{in}}} }^{r_{{\rm{out}}} } {\int\limits_0^H {\frac{{dH'}}{{{\textstyle{{dr} \over {dt}}}}}} } dr .\label{eq:10}
\end{eqnarray}
The above expression is due to the Hamilton equation $dr/dt = dH/dp_r |_r 
$ where $r$ and $p_r$ are canonical variables (in this case, the radial component of the radius and the momentum). As a reminder, the action of a tunneled particle in a potential barrier higher than the energy of the particle itself will be imaginary, $p_r  = \sqrt {2m\left( {E - V} \right)} 
$. Different from discussions of several authors for a static black hole mass [e.g. \cite{3}, \cite{5}, \cite{6}, \cite{7}, \cite{8}, \cite{12}, \cite{13}], the outgoing particle's energy must be time dependent for black holes with varying mass. So, the $dH'
$ integration at (\ref{eq:10}) is for all values of outgoing particle's energy, say from zero to $+ E\left( t \right)
$.\newline
By using approximation (\ref{eq:9}), we can perform the integration (\ref{eq:10}). For $dr$ integration, we can perform a contour integration for upper half complex plane to avoid the coordinate singularity $r_h$. The result is
\begin{eqnarray}
	{\mathop{\rm Im}\nolimits} S = \frac{{2 \pi E\left( t \right)}}{{\sqrt {F'\left( {r_h ,t} \right)G'\left( {r_h ,t} \right)} }} .\label{eq:11}
\end{eqnarray}
Since the tunneling probability is given by $\Gamma  \sim \exp \left[ { - {\textstyle{2 \over \hbar }}{\mathop{\rm Im}\nolimits} S} \right]
$, equalizing it with the Boltzmann factor $\exp \left[ { - \beta E\left( t \right)} \right]
$ for a system with time dependent energy we obtain 
\begin{eqnarray}
	T_H  = \frac{{\hbar \sqrt {F'\left( {r_h ,t} \right)G'\left( {r_h ,t} \right)} }}{{4\pi }} .\label{eq:12}
\end{eqnarray}
As seen in expression (\ref{eq:12}), the Hawking temperature $T_H$ is time dependent. We will see it later that it is also radius dependent. Inserting the values of $F'\left( {r_h ,t} \right)
$ and $G'\left( {r_h ,t} \right)
$ with the condition of $\left( {m'} \right)^2  \simeq 0
$, one has
\begin{eqnarray}
	T_H  = \frac{{\hbar \sqrt {\left( {m\left( {r_h ,t} \right)} \right)^2  - 2m'\left( {r_h ,t} \right)\left( {m\left( {r_h ,t} \right) + 2m\left( {r_h ,t} \right)} \right)} }}{{8\pi \left( {m\left( {r_h ,t} \right)} \right)^2 }} .\label{eq:13}
\end{eqnarray}
In the above, the mass $m$ is written as $m\left( {r_h ,t} \right)
$ explicitly to remind us that the mass and its  derivative $m'\left( {r_h ,t} \right)
$ are evaluated at $r_h$ (the radius of event horizon). Interestingly, for a black hole whose mass is only time dependent, not radius dependent, $m' = 0
$, we  get
\begin{eqnarray}
	T_H  = \frac{\hbar }{{8\pi m\left( t \right)}} .\label{eq:14}
\end{eqnarray}
and thus, we recover to the time-dependent Hawking temperature. Further interesting investigation would be to understand the black hole model with $x(m) =  - C(m)/m
$, however it will not be discussed in this paper.

\section{Hawking Temperature from Hamilton-Jacobi Method in Vaidya Space-Time}
In this section, we would work in scalar field theory with gravitational background. We still work in the metric (\ref{eq:1}) and impose the condition $\dot m = {{ - C\left( m \right)} \mathord{\left/
 {\vphantom {{ - C\left( m \right)} {m^2 }}} \right.
 \kern-\nulldelimiterspace} {m^2 }}
$ later. In this section, the condition $\left( {\dot m} \right)^2  \simeq 0
$ and $\left( {x\left( m \right)} \right)^2  \simeq 0
$ are very important in showing the proportionality between each expansion terms of action. To avoid confusion with $F\left( {r,t} \right)
$ and $G\left( {r,t} \right)
$ that have been used in the previous section, we rewrite metric (\ref{eq:1}) as
\begin{eqnarray}
	ds^2  =  - f\left( {r,t} \right)dt^2  + g\left( {r,t} \right)^{ - 1} dr^2  + r^2 d\Omega ^2 ,\label{eq:15}
\end{eqnarray}
where $f\left( {r,t} \right) \equiv \dot m(1 - 2mr^{ - 1} )/x(m)
$ and $g\left( {r,t} \right) \equiv \left( {1 - 2mr^{ - 1} } \right)
$.\newline
Massless scalar particles under the gravitational background $g_{\mu \nu } 
$ obey the Klein-Gordon equation
\begin{eqnarray}
	\frac{{ - \hbar ^2 }}{{\sqrt { - g} }}\partial _\mu  \left[ {g^{\mu \nu } \sqrt { - g} \partial _\nu  } \right]\phi  = 0 .\label{eq:16}
\end{eqnarray}
For spherical symmetric black hole, we may reduce our attention only to ($r-t$) sector in the space-time, or in other words, we reduce to two dimensional black hole problems. In this consideration, equation (\ref{eq:16}) under the background metric (\ref{eq:15}) simplifies to
\begin{eqnarray}
	\partial _t ^2 \phi  - \frac{1}{{2fg}}\left( {\dot fg + \dot gf} \right)\partial _t \phi  - \frac{1}{2}\left( {f'g + fg'} \right)\partial _r \phi  - fg\partial _r ^2 \phi  = 0 .\label{eq:17}
\end{eqnarray}
In the above, $f = f\left( {r,t} \right)
$ and $g = g\left( {r,t} \right)
$. By the standard ansatz for scalar wave function $\phi \left( {r,t} \right) = \exp \left[ { - {\textstyle{i \over \hbar }}S\left( {r,t} \right)} \right]
$, equation (\ref{eq:17}) leads to the equation for the action $S\left( {r,t} \right)
$
\begin{eqnarray}
	\begin{array}{l}
 \left( {\frac{{ - i}}{\hbar }\left( {\frac{{\partial ^2 S}}{{\partial t^2 }}} \right)} \right) - \frac{1}{{\hbar ^2 }}\left( {\frac{{\partial S}}{{\partial t}}} \right)^2  - \frac{1}{{2fg}}\left( {\dot fg + \dot gf} \right)\left( {\frac{{ - i}}{\hbar }} \right)\left( {\frac{{\partial S}}{{\partial t}}} \right) \\ 
  - \frac{1}{2}\left( {f'g + fg'} \right)\left( {\frac{{ - i}}{\hbar }} \right)\left( {\frac{{\partial S}}{{\partial r}}} \right) - fg\left( {\frac{{ - i}}{\hbar }\left( {\frac{{\partial ^2 S}}{{\partial r^2 }}} \right) - \frac{1}{{\hbar ^2 }}\left( {\frac{{\partial S}}{{\partial r}}} \right)^2 } \right) = 0 \\ 
 \end{array} .\label{eq:18}
\end{eqnarray}
Now, our next step is to solve this equation. An approximation method can be applied by expanding the action in the order of Planck constant power,
\begin{eqnarray}
	S\left( {r,t} \right) = S_0 \left( {r,t} \right) + \sum\nolimits_n {\alpha _n \hbar ^n S_n \left( {r,t} \right)} ,\label{eq:19}
\end{eqnarray}
for $n=1,2,3,...$. The constant $\alpha _n 
$ is set to keep all the expansion terms have the action's dimension. Taking unit dimensions $G = c = k_B  = 1
$, $\alpha _n 
$ would have the dimension of $[m]^{ - 2n} 
$ which $m$ refers to the mass. It is clear that this expansion would lead to a very long equation. Due to the very small value of the Planck constant, many authors \cite{5}, \cite{6}, \cite{12}, \cite{13} neglect the terms for $n \ge 1
$. This consideration is acceptable, and including higher terms is just adding correction for semi-classical derivation of Hawking temperature. By grouping all the terms into the same powers of $\hbar 
$, we could write for some lowest rank as
\begin{eqnarray}	\hbar ^0 {\rm{  }}:{\rm{   }}fg\left( {\partial _r S_0 } \right)^2  - \left( {\partial _t S_0 } \right)^2  = 0 ,\label{eq:20}
\end{eqnarray}
\[
\begin{array}{l}
 \hbar ^1 {\rm{  }}:{\rm{   }}ifg\partial _r ^2 S_0  - i\left( {\partial _t ^2 S_0 } \right) + 2fg\left( {\partial _r S_0 } \right)\left( {\partial _r S_1 } \right) + \frac{i}{2}\left( {f'g + fg'} \right)\partial _t S_0  \\ 
 {\rm{           }} + \frac{i}{2}\frac{{\left( {\dot fg + \dot gf} \right)}}{{fg}}\partial _t S_0  - 2\left( {\partial _t S_0 } \right)\left( {\partial _t S_1 } \right) = 0 ,\\ 
 \end{array}
\]
\[
\begin{array}{l}
 \hbar ^2 {\rm{  }}:{\rm{   }}ifg\partial _r ^2 S_1  - i\partial _t ^2 S_1  + 2fg\left( {\partial _r S_0 } \right)\left( {\partial _r S_2 } \right) + fg\left( {\partial _r S_1 } \right)^2  + \frac{i}{2}\left( {f'g + fg'} \right)\partial _t S_1  \\ 
 {\rm{           }} + \frac{i}{2}\frac{{\left( {\dot fg + \dot gf} \right)}}{{fg}}\partial _t S_1  - 2\left( {\partial _t S_0 } \right)\left( {\partial _t S_2 } \right) - \left( {\partial _t S_1 } \right)^2  = 0 ,\\ 
 \end{array}
\]
\[
\begin{array}{l}
 \hbar ^3 {\rm{  }}:{\rm{   }}ifg\partial _r ^2 S_2  - i\partial _t ^2 S_2  + 2fg\left( {\partial _r S_0 } \right)\left( {\partial _r S_3 } \right) +  + 2fg\left( {\partial _r S_1 } \right)\left( {\partial _r S_2 } \right) - 2\left( {\partial _t S_0 } \right)\left( {\partial _t S_3 } \right) \\ 
 {\rm{          }} - 2\left( {\partial _t S_1 } \right)\left( {\partial _t S_2 } \right) + \frac{i}{2}\left( {f'g + fg'} \right)\partial _t S_2  + \frac{i}{2}\frac{{\left( {\dot fg + \dot gf} \right)}}{{fg}}\partial _t S_2  = 0 ,\\ 
 \end{array}
\]
\[...
\]
\[...
\]
As obtained by Banerjee and Majhi \cite{8} for the metric that has a time-like Killing vector, the metric (\ref{eq:15}) also  leads to such a relation:
\[
\hbar ^0 {\rm{  }}:{\rm{   }}\partial _t S_0  =  \pm \sqrt {fg} \partial _r S_0 ,
\]
\begin{eqnarray}
\hbar ^1 {\rm{  }}:{\rm{   }}\partial _t S_1  =  \pm \sqrt {fg} \partial _r S_1 ,\label{eq:21}
\end{eqnarray}
\[
\hbar ^2 {\rm{  }}:{\rm{   }}\partial _t S_2  =  \pm \sqrt {fg} \partial _r S_2 ,
\]
\[...
\]
\[...
\]
To get the benefits of the conditions for black holes with slowly varying mass, $(m')^2  \simeq 0
$ and $(\dot m)^2  \simeq 0
$, the arbitrary function  $x\left( m \right)
$ must be taken to be $-m'$ \cite{9}. This mechanism means that we have used the dynamic equation for black hole mass as $
m' = C\left( m \right)/m$. Unless these conditions, the set of equations (\ref{eq:21}) can not be obtained. Then one can obtain a general pattern for arbitrary $n$, thus for the terms of $\hbar ^n 
$, that is $\partial _t S_n  =  \pm \sqrt {fg} \partial _r S_n 
$. From the set of equations (\ref{eq:20}), one can see that each $S_n \left( {r,t} \right)
$ is proportional to $S_0 \left( {r,t} \right)
$. By this evidence, we can write the expansion (\ref{eq:19}) into 
\begin{eqnarray}
	S\left( {r,t} \right) = S_0 \left( {r,t} \right) + \sum\nolimits_n {\alpha _n \hbar ^n S_0 \left( {r,t} \right)}  = \left( {1 + \sum\nolimits_n {\alpha _n \hbar ^n } } \right)S_0 \left( {r,t} \right) .\label{eq:22}
\end{eqnarray}
The extra value $\sum\nolimits_n {\alpha _n \hbar ^n S_0 \left( {r,t} \right)} 
$ in (\ref{eq:22}) can be regarded as the correction term of the semi-classical analysis. So, our next step is to find the solution of $S_0 \left( {r,t} \right)
$ satisfying $\partial _t S_0  =  \pm \sqrt {fg} \partial _r S_0 
$.\newline
In the standard Hamilton-Jacobi method, $S_0 \left( {r,t} \right)
$ can be written into two parts, the time part which has the form of $Et$ and the radius part $\tilde S_0 \left( r \right)
$ which is in general a radius dependent only. Since our metric coefficients are both radius and time dependent, the standard method would not be applicable. We could generalized the method by making an ansatz
\begin{eqnarray}
	S_0 \left( {r,t} \right) = \int\limits_0^t {E\left( {t'} \right)dt'}  + \tilde S_0 \left( {r,t} \right) .\label{eq:23}
\end{eqnarray}
At the first sight, it seems that the ansatz is rather strange, that is $S\left( {r,t} \right)
$ and $\tilde S_0 \left( {r,t} \right)
$ are both $t$ and $r$ dependent. The term $\int {E\left( {t'} \right)dt'} 
$ is more understandable, since the emitted particle's energy is continuum and time dependent. Let see how it works.\newline
From (\ref{eq:23}), one can write that
\begin{eqnarray}
	\partial _t S_0 \left( {r,t} \right) = E\left( t \right) + \partial _t \tilde S_0 \left( {r,t} \right) \label{eq:24}
\end{eqnarray}
and
\begin{eqnarray}
	\partial _r S_0 \left( {r,t} \right) = \partial _r \tilde S_0 \left( {r,t} \right) .\label{eq:25}
\end{eqnarray}
Since $\tilde S_0 \left( {r,t} \right)
$ is $t$ and $r$ dependent, one can write
\begin{eqnarray}
	\frac{{d\tilde S_0 \left( {r,t} \right)}}{{dr}} = \partial _r \tilde S_0 \left( {r,t} \right) + \partial _t \tilde S_0 \left( {r,t} \right)\frac{{dt}}{{dr}} .\label{eq:26}
\end{eqnarray}
Eliminating $dt/dr$ by the use of ${{dr} \mathord{\left/
 {\vphantom {{dr} {dt}}} \right.
 \kern-\nulldelimiterspace} {dt}} =  \pm \sqrt {fg} 
$, equation (\ref{eq:26}) can be written as
\begin{eqnarray}
	\frac{{d\tilde S_0 \left( {r,t} \right)}}{{dr}} \mp \frac{1}{{\sqrt {fg} }}\partial _t \tilde S_0 \left( {r,t} \right) = \partial _r \tilde S_0 \left( {r,t} \right) .\label{eq:27}
\end{eqnarray}
Combining the first equation of (\ref{eq:21}), with equations (\ref{eq:24}) and (\ref{eq:27}) where we should note that the action equation $- \left( {fg} \right)^{ - 1/2} \partial _t S_0 \left( {r,t} \right) = \partial _r S_0 \left( {r,t} \right)
$ is belong to outgoing particle and with the radial evolution is $dr/dt = \sqrt {fg} 
$, then we could write
\begin{eqnarray}
	\mp \left( {fg} \right)^{ - 1/2} \left( {E\left( t \right) + \partial _t \tilde S_0 \left( {r,t} \right)} \right) = \frac{{d\tilde S_0 \left( {r,t} \right)}}{{dr}} \mp \left( {fg} \right)^{ - 1/2} \partial _t \tilde S_0 \left( {r,t} \right) .\label{eq:28}
\end{eqnarray}
From (\ref{eq:28}), we can get the exact differentiation of $\tilde S_0 \left( {r,t} \right)
$
\begin{eqnarray}
	\frac{{d\tilde S_0 \left( {r,t} \right)}}{{dr}} =  \mp \left( {fg} \right)^{ - 1/2} E\left( t \right),\label{eq:29}
\end{eqnarray}
and the solution of $\tilde S_0 \left( {r,t} \right)
$ can be obtained by integration
\begin{eqnarray}
	\tilde S_0 \left( {r,t} \right) =  \mp E\left( t \right)\int {\frac{{dr}}{{\sqrt {fg} }}} .\label{eq:30}
\end{eqnarray}
After inserting the mass evolution equation, $\dot m = -{{C\left( m \right)} \mathord{\left/
 {\vphantom {{C\left( m \right)} {m^2 }}} \right.
 \kern-\nulldelimiterspace} {m^2 }}
$, and the arbitrary function $x\left( m \right) = - {{C\left( m \right)} \mathord{\left/
 {\vphantom {{ C\left( m \right)} m}} \right.
 \kern-\nulldelimiterspace} m}
$, one may identify that $f$ and $g$ are exactly equal to $F$ and $G$ stated in the previous section. By this equality, the integration (\ref{eq:30}) can be evaluated by adopting the value of $\int {\left( {FG} \right)^{ - 1/2} dr} 
$ as in obtaining expression (\ref{eq:11}) from (\ref{eq:10}) along with it's approximation method (near horizon Taylor expansion). The result for the integration (\ref{eq:30}) is
\begin{eqnarray}
	\tilde S_0 \left( {r,t} \right) =  \mp E\left( t \right)\frac{{i\pi }}{{\sqrt {F'G'} }}.\label{eq:31}
\end{eqnarray}
The equation gives the complete action
\begin{eqnarray}
	S\left( {r,t} \right) = \left( {1 + \sum\nolimits_n {\alpha _n \hbar ^n } } \right)\left( {\int\limits_0^t {E\left( {t'} \right)dt'}  \mp E\left( t \right)\frac{{i\pi }}{{\sqrt {F'G'} }}} \right) .\label{eq:32}
\end{eqnarray}
The signs $+ \left(  -  \right)
$ in expression (\ref{eq:32}) refer to the action for ingoing (outgoing) particle.\newline 
Back to our first ansatz for scalar wave function, $\phi  = \exp \left[ {{\textstyle{{ - i} \over \hbar }}S\left( {r,t} \right)} \right]
$, the wave function for ingoing and outgoing massless scalar particle can be read of as
\begin{eqnarray}
	\phi _{in} \left( {r,t} \right) = \exp \left[ { - \frac{i}{\hbar }\left( {1 + \sum\nolimits_n {\alpha _n \hbar ^n } } \right)\left( {\int\limits_0^t {E\left( {t'} \right)dt'}  + E\left( t \right)\frac{{i\pi }}{{\sqrt {F'G'} }}} \right)} \right] \label{eq:33}
\end{eqnarray}
and
\begin{eqnarray}
	\phi _{out} \left( {r,t} \right) = \exp \left[ { - \frac{i}{\hbar }\left( {1 + \sum\nolimits_n {\alpha _n \hbar ^n } } \right)\left( {\int\limits_0^t {E\left( {t'} \right)dt'}  - E\left( t \right)\frac{{i\pi }}{{\sqrt {F'G'} }}} \right)} \right] \label{eq:34}
\end{eqnarray}
respectively. Consequently, from (\ref{eq:33}) one can get the ingoing probability of particle as below
\begin{eqnarray}
	P_{in}  = \exp \left[ {\frac{2}{\hbar }\left( {1 + \sum\nolimits_n {\alpha _n \hbar ^n } } \right)\left( {{\mathop{\rm Im}\nolimits} \int\limits_0^t {E\left( {t'} \right)dt'}  + \frac{{\pi E\left( t \right)}}{{\sqrt {F'G'} }}} \right)} \right] .\label{eq:36}
\end{eqnarray}
This ingoing probability must be equal to unity since all particles including the massless one are absorbed by the black hole. This consideration gives us the relation 
\[
{\mathop{\rm Im}\nolimits} \int\limits_0^t {E\left( {t'} \right)dt'}  =  - \frac{{\pi E\left( t \right)}}{{\sqrt {F'G'} }},
\]
which leads the outgoing probability
\begin{eqnarray}
	P_{out}  = \exp \left[ { - \frac{{4\pi E\left( t \right)}}{{\hbar \sqrt {F'G'} }}\left( {1 + \sum\nolimits_n {\alpha _n \hbar ^n } } \right)} \right] .\label{eq:37}
\end{eqnarray}
Finally, to get the Hawking temperature from the outgoing probability (\ref{eq:37}), we equate this probability expression with $\exp \left[ { - \beta E\left( t \right)} \right]
$ which in \cite{5} is called 'detailed balance' principle. It yields
\begin{eqnarray}
	T_{H'}  = \frac{{\hbar \sqrt {F'\left( {r_h ,t} \right)G'\left( {r_h ,t} \right)} }}{{4\pi \left( {1 + \sum\nolimits_n {\alpha _n \hbar ^n } } \right)}} .\label{eq:38}
\end{eqnarray}
We use the index $H'$ to distinguish the result (\ref{eq:38}) with that of (\ref{eq:12}). 
Neglecting the correction term $
\sum\nolimits_n {\alpha _n \hbar ^n } $ in the denominator of (\ref{eq:38}), one has recovered result (\ref{eq:12}) which has been obtained by the radial null geodesics at the previous section.

\section{Conclusion}
We have worked out the Hawking temperature for the metric with no time-like Killing vector and with the time and radius dependent coefficients. It has been shown that by the use of mass evolution $\partial m/\partial t =  - C(m)/m^2 
$ and mass dependent function $x(m) =  - C(m)/m
$, the  resulting Hawking temperature coincides with that for the widely known time dependent one in the case of radius independent mass. By neglecting the $
\sum\nolimits_n {\alpha _n \hbar ^n } $ term in the denominator of the resulting Hawking temperature for the case of beyond semi-classical approach as given in Section 3, one recovers the result in the section 2. However, the equal obtained results in (\ref{eq:12}) and (\ref{eq:37}) with neglected correction terms are actually based on some subtleties. In section 2, we use the $
\dot r
$ that is obtained from the coordinate after Painleve transformation which differs with a factor half with untransformed one (directly obtained from Schwarzschild like metric (\ref{eq:3})). This is clear since the particle moves in the path with no singularity\footnote{We thank B. Majhi and R. Banerjee for pointing this out.}. Thus, in section 3 we use the $
\dot r
$ in deriving the action's solution (\ref{eq:27}) that is obtained from Schwarzschild like metric (without Painleve transformation), since the equation of action comes from it (\ref{eq:17}). It needs further investigations to verify whether the results are still the same in the suitable Painleve coordinate in deriving the wave equation (\ref{eq:16}) (parallel to section 3.2 \cite{8}).
\newline
The slowly varying mass with respect both to time and radius, considered in Section 3, affects our result on the Hawking temperature expression and the proportionality between each terms of action's expansion. We have restored the $x(m)$ as $-m'$ in section 3 to get the proportionality by using our slowly varying mass condition. This consideration of course yields an equation for mass dynamics $
m' = C\left( m \right)/m $ which needs further investigations. Of course, our analysis is not valid for the case of highly radiated black holes. There are several models that have been proposed for describing black hole radiation. But, the lack of experimental data does not enable one to compare between the models.
\newline
Further investigation on the relationship between ingoing and outgoing probabilities for the case of black hole with varying mass/energy by the use of path integral method is intriguing. This work was first performed by Hartle and Hawking \cite{14} for a black hole with constant mass and constant energy of outgoing particle. The relation proposed in the paper, $P_{out}  = \exp \left[ { - \beta E\left( t \right)} \right]P_{in} 
$, is an intuitive manner as it might not be that simple.

\section{Acknowledgements}
HMS thanks R. Kurniadi and A. Nugroho from the Faculty of Mathematics and Natural Sciences, ITB for useful discussions. HMS also thanks B. Majhi and R. Banerjee from S. N. Bose National Centre for Basic Sciences for useful comments.


\begin{thebibliography}{99}

\bibitem{1} S. W. Hawking (1975), \textit{Particle Creation by Black Holes}, \textit{Comm. Math. Phys.} \textbf{43}, 199. \\
R. M. Wald (1994), \textit{Quantum Field Theory in Curved Spacetime and Black Hole Thermodynamics}, The Univ. of Chicago Press.

\bibitem{2} R. M. Wald (1984), \textit{General Relativity}, The Univ. of Chicago Press. 

\bibitem{3} M. K. Parikh and F. Wilczek (2000), \textit{Hawking Radiation as Tunneling}, \textit{Phys. Rev. Lett.} \textbf{85}, 5042 [arXiv:hep-th/9907001].

\bibitem{4} M. K. Parikh (1998), \textit{Membrane Horizons : The Black Hole's New Clothes}, Ph.D. Thesis, Princeton University, Princeton, NJ [arXiv:hep-th/9907002].

\bibitem{5} K. Srinivasan and T. Padmanabhan (1999), \textit{Particle Production and Complex Path Analysis}, \textit{Phys. Rev.} \textbf{D 60}, 024007 [arXiv:hep-th/9907002].\\
S. Shankaranarayanan, T. Padmanabhan, and K. Srinivasan (2002), \textit{Hawking Radiation in Different Coordinate Settings : Complex Path Approach}, \textit{Class. Quant. Grav.} \textbf{19}, 2671, [arXiv:gr-qc/0010042].

\bibitem{6} P. Mitra (2006), \textit{Hawking Temperature from Tunneling Formalism}, \textit{Phys. Lett.} \textbf{B 648}, 240 [arXiv:hep-th/0611265].

\bibitem{7} E.T. Akhmedov, V. Akhmedova, and D. Singleton (2006), \textit{Hawking Temperature in the Tunneling Picture}, \textit{Phys. Lett.} \textbf{B 642}, 124 [arXiv:hep-th/0608098].\\
E.T. Akhmedov, V. Akhmedova, D. Singleton, and T. Pilling (2007), \textit{Thermal Radiation of Various Gravitational Background}, \textit{Int. J. Mod. Phys.} \textbf{A 22}, 1705, [arXiv:hep-th/0605137].\\
T. Pilling (2008), \textit{Black Hole Thermodynamics and the Factor 2 Problem}, \textit{Phys. Lett.} \textbf{B 660}, 402, [arXiv:0709.1624].\\
B.D. Chowdhury (2008), \textit{Pramana} \textbf{70}, 593, [arXiv:hep-th/0605197].\\
A.J.M. Medved and Elias C. Vagenas (2005), \textit{On Hawking Radiation as Tunneling with Back-Reaction}, \textit{Mod. Phys. Lett.} \textbf{A 20} , 2449, [arXiv:gr-qc/0504113].\\
Ryan Kerner and Robert B. Mann (2006), \textit{Tunnelling, Temperature and Taub-NUT Black Holes}, \textit{Phys. Rev.} \textbf{D 73}, 104010, [arXiv:gr-qc/0603019].\\
Michele Arzano, A.J.M. Medved and E. C. Vagenas (2005), \textit{Hawking Radiation as Tunneling through the Quantum Horizon}, \textit{JHEP} \textbf{0509}, 037, [arXiv:hep-th/0505266].\\
Rabin Banerjee and Bibhas R. Majhi (2008), \textit{Quantum Tunneling and Back Reaction}, \textit{Phys.Lett.} \textbf{B 662}, 62, [arXiv:0801.0200].\\
Rabin Banerjee, Bibhas R. Majhi, and Saurav Samanta (2008), \textit{Noncommutative Black Hole Thermodynamics}, \textit{Phys.Rev.} \textbf{D 77}, 124035, [arXiv:0801.3583].

\bibitem{8} R. Banerjee and Bibhas R. Majhi (2008), \textit{Quantum Tunneling Beyond Semiclassical Approximation}, \textit{JHEP} \textbf{0806}, 095, [arXiv:0805.2220].

\bibitem{9} A. N. St. J. Farley and P. D. D'Eath (2006), \textit{Vaidya Space-Time in Black Hole Evaporation}, \textit{Gen. Rel. Grav.} \textbf{38}, 425, [arXiv:gr-qc/0510040].

\bibitem{10} S.W. Hawking (1974), \textit{Nature}, \textbf{248}, 30.

\bibitem{11} S. Massar (1995), \textit{The Semi-Classical Back Reaction to Black Hole Evaporation}, \textit{Phys. Rev.} \textbf{D 52}, 5857, [arXiv:gr-qc/9411039].

\bibitem{12} Marco Angheben, Mario Nadalini, Luciano Vanzo, and Sergio Zerbini (2005), \textit{Hawking Radiation as Tunneling for Extremal and Rotating Black Holes}, \textit{JHEP} \textbf{0505}, 014, [arXiv:hep-th/0503081].\\
M. Nadalini, L. Vanzo and S. Zerbini (2005), \textit{Hawking radiation as tunneling: The D dimensional rotating case}, \textit{J. Phys.} \textbf{A 39}, 6601 [arXiv:hep-th/0511250].


\bibitem{13} E.T. Akhmedov, T. Pilling, and Douglas Singleton, \textit{Subtleties in the Quasi-Classical Calculation of Hawking Radiation}, [arXiv:0805.2653].

\bibitem{14} J.B. Hartle and S.W. Hawking (1976), \textit{Path Integral Derivation of Black Hole Radiance}, \textit{Phys. Rev.} \textbf{D 13}, 2188.

\bibitem{15} Sujoy K. Modak (2008), \textit{Corrected Entropy of BTZ Black Hole in Tunneling Approach}, [arXiv:0807.0959].\\
Bibhas R. Majhi (2008), \textit{Fermion Tunneling Beyond Semiclassical Approximation}, [arXiv:0809.1508].\\
Rabin Banerjee and Bibhas R. Majhi (2008), \textit{Quantum Tunneling and Trace Anomaly}, [arXiv:0808.3688].

\bibitem{16} E.C. Vagenas (2000), \textit{Are extremal 2-D black holes really frozen?}, \textit{Phys. Lett.} \textbf{B 503}, 399 [arXiv:hep-th/0012134].\\
E.C. Vagenas (2001), \textit{Two-dimensional dilatonic black holes and Hawking radiation}, \textit{Mod. Phys. Lett.} \textbf{A 17}, 609 [arXiv:hep-th/0108147].\\
E.C. Vagenas (2001), \textit{Semiclassical corrections to the Bekenstein-Hawking entropy of the BTZ black hole via selfgravitation}, \textit{Phys. Lett.} \textbf{B 533}, 302 [arXiv:hep-th/0109108].\\
A. J. M. Medved and E. C. Vagenas (2005), \textit{On Hawking radiation as tunneling with logarithmic corrections}, \textit{Mod. Phys. Lett.} \textbf{A 20}, 1723 [arXiv:gr-qc/0505015].\\
Qing-Quan Jiang and Shuang-Qing Wu (2005), \textit{Hawking radiation of charged particles as
tunneling from Reissner-Nordstrom-de Sitter black holes with a global monopole}, \textit{Phys. Lett.} \textbf{B 635}, 151 [arXiv:hep-th/0511123].\\
Qing-Quan Jiang, Shuang-Qing Wu and Xu Cai (2005), \textit{Hawking radiation as tunneling from
the Kerr and Kerr-Newman black holes}, \textit{Phys. Rev.} \textbf{D 73} 064003
[arXiv:hep-th/0512351].\\
R. Di Criscienzo, M. Nadalini, L. Vanzo, S. Zerbini and G. Zoccatelli (2007), \textit{On the Hawking radiation as tunneling for a class of dynamical black holes}, \textit{Phys. Lett.} \textbf{B 657}, 107 [arXiv:0707.4425].

\bibitem{17} E.C. Vagenas (2001), \textit{Complex paths and covariance of Hawking radiation in 2-D stringy black holes}, \textit{Nuovo Cim.} \textbf{B 117}, 899 [hep-th/0111047].\\

\bibitem{18} S.A. Hayward, R. Di Criscienzo, L. Vanzo, M. Nadalini and S. Zerbini (2008), \textit{Local Hawking temperature for dynamical black holes}, [arXiv:0806.0014].

\end{thebibliography}
\end{document}